# Testing of planar hydrogenated amorphous silicon sensors with charge selective contacts for the construction of 3D- detectors.

M. Menichelli[1,*], M. Bizzarri[1,2], M. Boscardin[3,4], M. Caprai[1], A.P. Caricato[5], G.A.P. Cirrone[6], M. Crivellari[4], I. Cupparo[7], G. Cuttone[6], S. Dunand[8], L. Fanò[1,2], O. Hammad[4], M. Ionica[1], K. Kanxheri[1], M. Large[10], G. Maruccio[5], A.G. Monteduro[5], A. Morozzi[1], F. Moscatelli[1,9], A. Papi[1], D. Passeri[1,11], M. Petasecca[10], G. Petringa[6], G. Quarta[5], S. Rizzato[5], A. Rossi[1,2], G. Rossi[1], A. Scorzoni[1,11], L. Servoli[1], C. Talamonti[7], G. Verzellesi[3,12], N. Wyrsch[8].

[1.] INFN, Sez. di Perugia, via Pascoli s.n.c. 06123 Perugia (ITALY)
[2.] Dip. di Fisica e Geologia dell'Università degli Studi di Perugia, via Pascoli s.n.c. 06123 Perugia (ITALY)
[3.] INFN, TIPFA Via Sommarive 14, 38123 Povo (TN) (ITALY)
[4.] Fondazione Bruno Kessler, Via Sommarive 18, 38123 Povo (TN) (ITALY)
[5.] INFN and Department of Mathematics and Physics "Ennio de Giorgi" University of Salento, Via per Arnesano, 73100 Lecce (ITALY)
[6.] INFN Laboratori Nazionali del Sud, Via S.Sofia 62, 95123 Catania (ITALY)
[7.] INFN and Dipartimento di Fisica Scienze Biomediche sperimentali e Cliniche "Mario Serio", Viale Morgagni 50, 50135 Firenze (FI) (ITALY)
[8.] Ecole Polytechnique Fédérale de Lausanne (EPFL), Photovoltaic and Thin-Film Electronics Laboratory (PV-Lab), Rue de la Maladière 71b, 2000 Neuchâtel, (SWITZERLAND).
[9.] CNR-IOM, via Pascoli s.n.c. 06123 Perugia (ITALY)
[10.] Centre for Medical Radiation Physics, University of Wollongong, Northfields Ave Wollongong NSW 2522, (AUSTRALIA).
[11.] Dip. di Ingegneria dell'Università degli studi di Perugia, via G.Duranti 06125 Perugia (ITALY)
[12.] Dip. di Scienze e Metodi dell'Ingegneria, Università di Modena e Reggio Emilia, Via Amendola 2, 42122 Reggio Emilia (ITALY)
* Corresponding author, Mauro.menichelli@pg.infn.it

# Abstract

Hydrogenated Amorphous Silicon (a-Si:H) is a well known material for its intrinsic radiation hardness and is primarily utilized in solar cells as well as for particle detection and dosimetry. Planar p-i-n diode detectors are fabricated entirely by means of intrinsic and doped PECVD of a mixture of Silane ($SiH_4$) and molecular Hydrogen. In order to develop 3D detector geometries using a-Si:H, two options for the junction fabrication have been considered: ion implantation and charge selective contacts through atomic layer deposition. In order to test the functionality of the charge selective contact electrodes, planar detectors have been fabricated utilizing this technique. In this paper, we provide a general overview of the 3D fabrication project followed by the results of leakage current measurements and x-ray dosimetric tests performed on planar diodes containing charge selective contacts to investigate the feasibility of the charge selective contact methodology for integration with the proposed 3D detector architectures.

## 1. Introduction

Hydrogenated amorphous silicon (a-Si:H) has remarkable radiation resistance properties and can be deposited on a variety of different substrates. The motivations for the construction of a 3D a-Si:H detector are described in ref. [1]. During 3D detector fabrication, once the a-Si:H substrate is grown, holes (or trenches) should be made in the substrate in order to prepare cavities for electrode manufacturing. Once the holes are etched their walls should then be doped in order to build the basic p-i-n electrode structure of the detector. Since commonly used techniques for planar structures (i.e. PECVD deposition of doped a-Si:H) are unable to achieve a conformal coating of the walls of the etched holes, two alternative options will be considered. The first option is the Atomic Layer Deposition (ALD) of conductive metallic oxides for the creation of selective contacts for each type of charge carrier. Titanium or aluminum doped zinc oxide (AZO) will be used as an electron selective contact and molybdenum oxide will be used for hole collection. This technique has been adopted in the construction of solar cells and, for the first time, is documented here in its use for detector fabrication. This charge collection mechanism is based on the different mobility of holes and electrons within the two sets of selective contact materials previously identified. Selective contact materials enhance the carrier mobility for electrons in the electron selective contact and for holes in the hole selective contact [2]. The second option is the Ion implantation of Phosphor ions for n-type doping and Boron for p-type doping.

In this paper we will describe the testing of planar diodes utilizing charge selective contacts to make a first assessment of their performance as radiation detectors. Specifically, the leakage current versus bias voltage for various samples and configurations (different thicknesses and charge selective contact materials) and the linearity of response to different x-ray dose rates will be measured. From this measurement, the charge sensitivity, a relevant parameters for x-ray dosimetry, will be estimated both at 0V, and under an applied bias.

## 2. Description of the samples and leakage current measurements

In order to assess the performance of the charge selective contact devices for particle detection and dosimetry, two sets of devices deposited on glass have been fabricated according to the scheme shown in figure 1a, while figure 1b shows a top view of the detector.

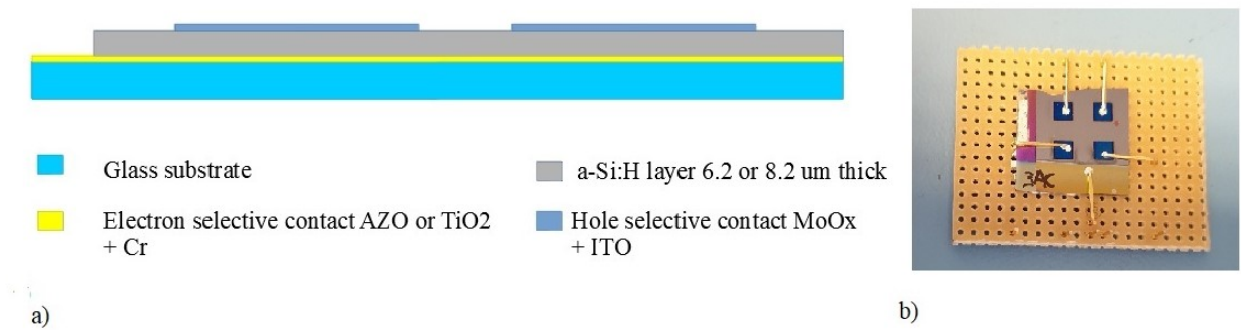

Fig.1 a) Layout of the detector (lateral view) b) Top view of the detector.

The detector layout consists of electron selective contacts in the form of ZnO:Al (Aluminum doped Zinc Oxide, AZO) or $TiO_2$ layers deposited on glass, with thicknesses of 60 nm and 10 nm, respectively. The active detector layer is made of a-Si:H in two different thickness: 6.2 and 8.2 microns. Hole selective contacts were made of 20 nm thick $MoO_x$ protected by a 60 nm thick indium tin oxide (ITO) layer. The hole selective contacts were deposited via sputtering resulting in four square electrodes having a 5 × 5 mm$^2$ area. For initial leakage current measurements, the detectors were biased using a Keithley 2410-C SMU to an electric field of up to 5 V/μm. Fig. 2 displays the resulting leakage currents versus bias voltage for the tested devices. During this measurement, the detectors were hosted in a climatic chamber that stabilizes the temperature at 21°C, the error bars shown in fig.2 are the RMS of the fluctuation around the average value during a continuous data taking lasting about 500 seconds . Figure 3a shows the time profile of current measured (row data) and figure 3b shows its distribution.

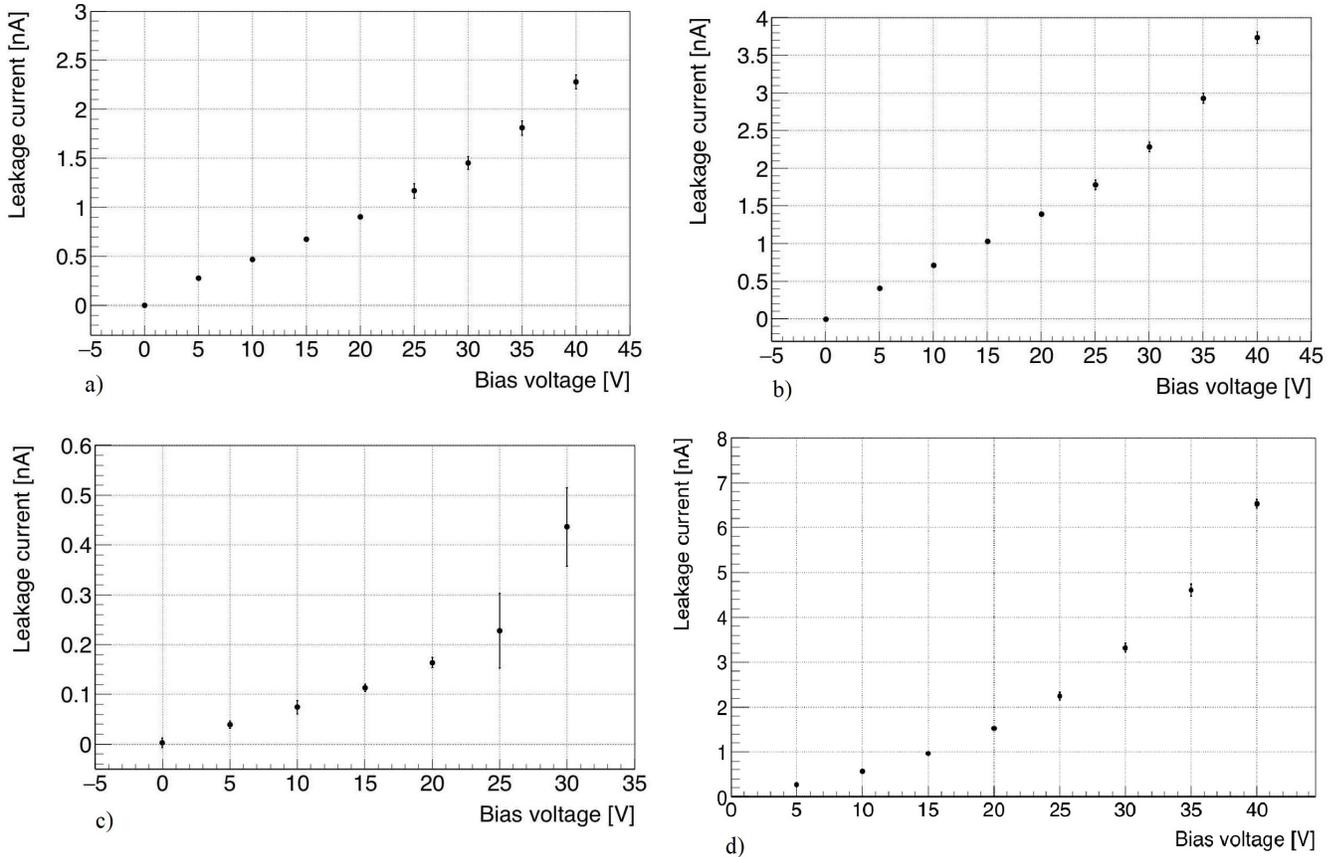

Fig. 2 Leakage current measurements of various devices in a-Si:H layer having different thicknesses and electron selective contact materials. a) AZO electron selective contacts with a-Si:H layer having 8.2 μm thickness (Sample 1). b) AZO electron selective contacts with a-Si:H layer having 8.2 μm thickness (Sample 2). c) AZO electron selective contacts with a-Si:H layer having 6.2 μm thickness. d) Titanium Oxide electron selective contacts with a-Si:H layer having 8.2 μm thickness.

From the data presented in Fig. 2 it is observed that AZO devices have a lower leakage current compared to the TiO$_2$ device, and that thinner devices at the same field have a lower current in comparison to their thicker counterparts in agreement with what reported in ref. [3] for doped contacts. The overall value of the leakage current normalized to the area for an 8.2 μm device with AZO selective contact at 5 V/μm bias is in the order of 9.2 nA/cm$^2$ for sample 1 and 14.8 nA/cm$^2$ for sample 2, while for the TiO$_2$ device is 24.4 nA/cm$^2$. For the AZO device having thickness 6.2 μm the leakage current per unit area at the same bias conditions is 1.76 nA/cm$^2$. Results are summarized in Table 1.

| Description of sample | Current density @ 30 V (nA/cm$^2$) | Current density @ 40 V (nA/cm$^2$) |
|---|---|---|
| 8.2 µm AZO electron selective contact. Sample 1 | 5.8 | 9.2 |
| 8.2 µm AZO electron selective contact. Sample 2 | 9.2 | 14.8 |
| 6.2 µm AZO electron selective contact. | 1.76 | N/A |
| 8.2 µm TiO$_2$ electron selective contact. | 12.8 | 24.4 |

Table 1. Current density for the 4 samples at 30 V and 40 V bias.

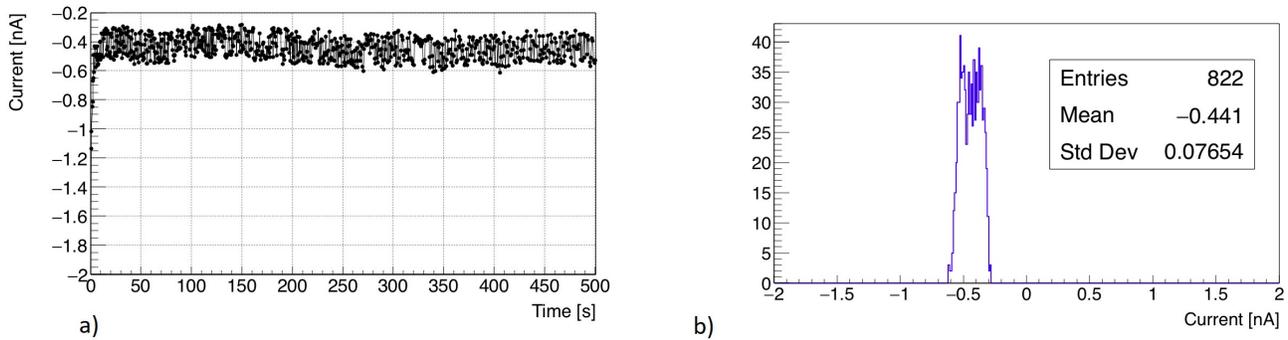

Fig.3 a) Time distribution of leakage current measurement at 30V on 6.2 µm AZO detector at 30V bias. b) measured leakage current values histogram.

Leakage current at various temperatures has also been measured on a 8.2 µm thick TiO$_2$ diode and also on a 8.2 µm thick AZO diode. The results are shown respectively in Fig. 4 and Fig.5. Fig 4a and Fig.5a show the current versus temperature at various bias voltages for the two devices while Fig. 4b and Fig. 5b show the same data displayed as current versus voltage at various temperatures. From the data we can observe that below 10 °C the current become lower than 100 pA but quite unstable.
 The leakage current versus temperature at various voltage follows quite well an exponential behavior especially on TiO$_2$ diode while in the AZO device the temperature slope is larger at the various voltages.

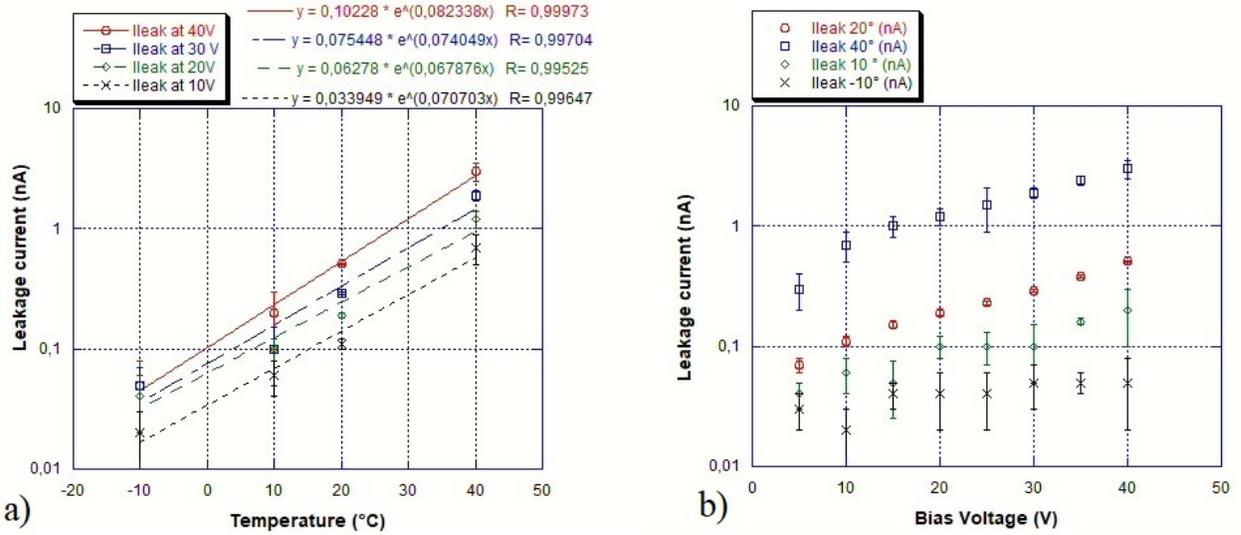

Fig.4 a) Leakage current versus temperature at various bias voltages on a 8.2 µm Titanium oxide device; an exponential fit is superposed on the data b) Leakage current versus bias voltage at various temperatures.

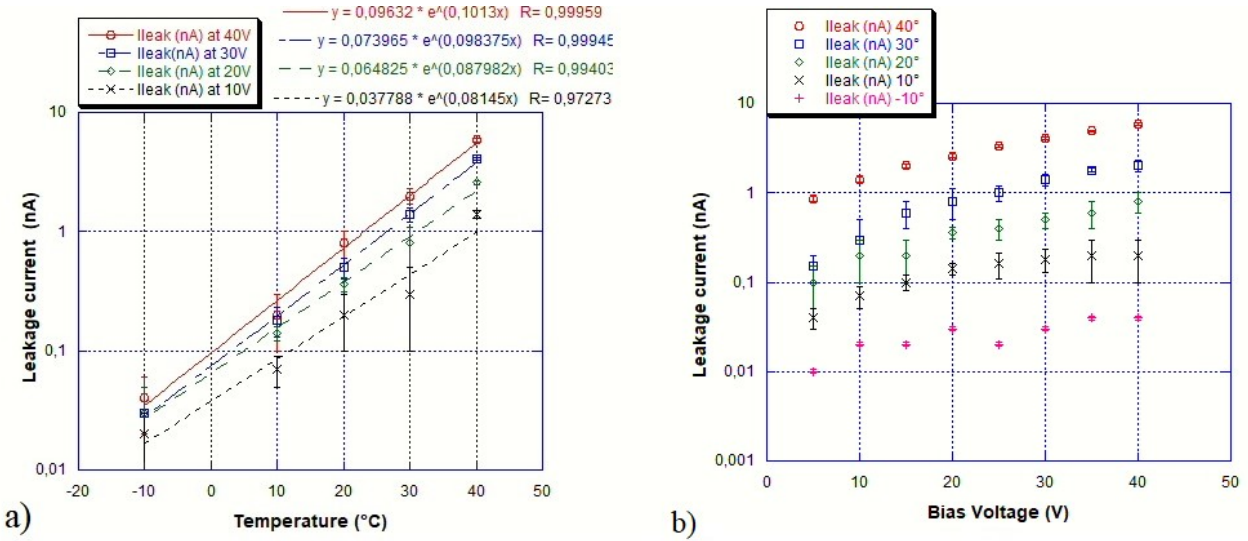

Fig.5 a) Leakage current versus temperature at various bias voltages on a 8.2 µm AZO device; an exponential fit is superposed on the data b) Leakage current versus bias voltage at various temperatures.

## 3. Current versus X-ray doses response in charge selective contact devices.

For preliminary determinations of the performance of the charge selective contact devices for dosimetry, the diodes where irradiated using a miniature X-ray tube from Newton Scientific [4] having 50 kV maximum voltage, 200 µA maximum current (10 W maximum power), and a 125 µm Be window. The irradiation setup is shown in Fig. 6.

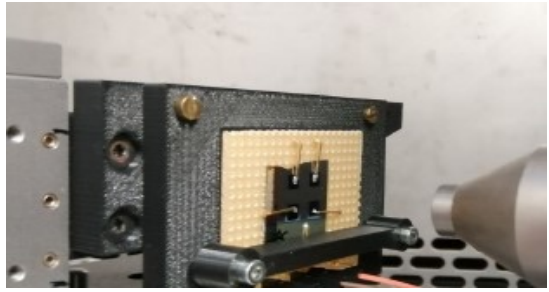

Fig.6 Irradiation test setup.

The X-ray tube was configured to deliver dose rates in the range of 0-22 mGy/s, with the response of the device calculated as the induced current to a beam exposure minus leakage current (at 0 Gy dose). For each of the device architectures investigated two measurements were performed: a first set of measurements with devices at 0V bias (photovoltaic mode) and a second set of measurements with devices under a non-zero applied bias (30 V for samples with 8.2 µm active layer thickness and 20V for the 6.2 µm thick sample). The results of measurements conducted at 0V bias are indicative of the devices performance for personal dosimetry in clinical applications, whilst their operation under a non-zero bias allows for the enhancement of the device sensitivities to assess their operation in the framework of the proposed 3D detector geometries as well as for beam monitoring applications. The test is performed by changing the tube current and measuring the device under test (DUT) current. Before the measurement a calibration of dose rate versus tube current in the same position of the detector is performed.

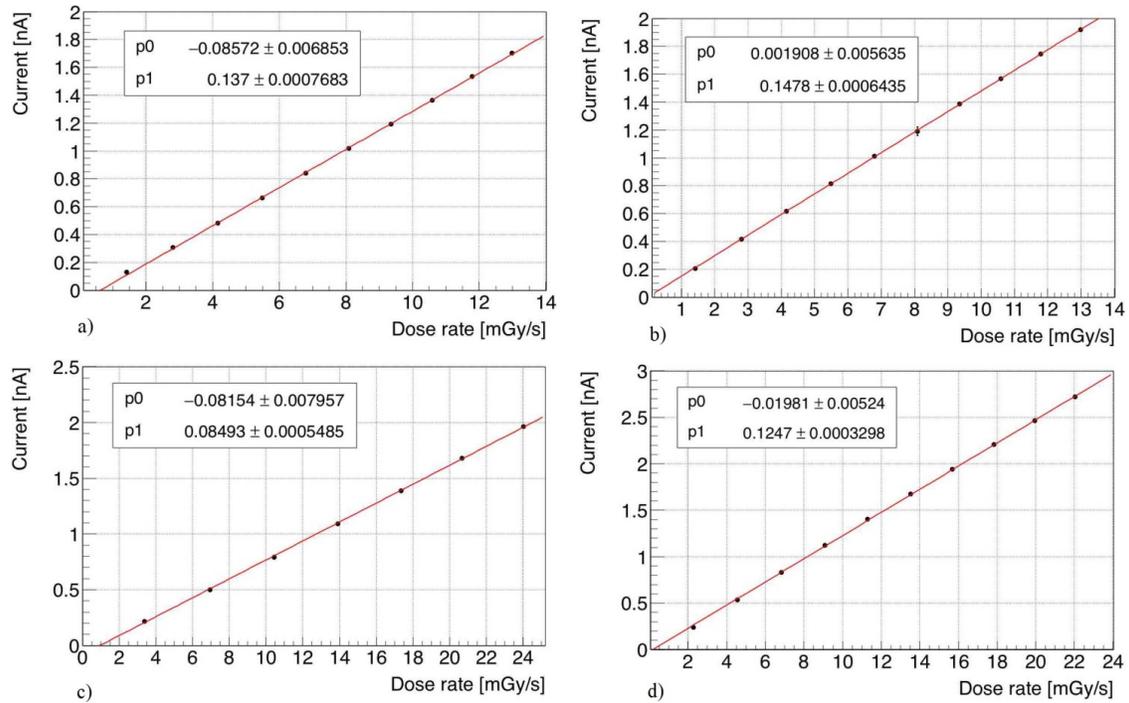

Fig.7 Current versus x-ray doses at 0V bias of various devices in a-Si:H having different layer thickness and electron selective contact materials a) AZO electron selective contacts with a-Si:H layer having 8.2 µm thickness (Sample 1) b) AZO electron selective contacts with a-Si:H layer having 8.2 µm thickness (Sample 2) c) AZO electron selective contacts with a-Si:H layer having 6.2 µm thickness d) Titanium Oxide electron selective contacts with a-Si:H layer having 8.2 µm thickness.

In figure 7 the results of the set of measurements at 0 V bias are shown, Table 2 displays the device sensitivities as calculated from the gradient of the dose linearity measurements presented in Figure 7. Also given in Table 2 is the regression coefficient of the linear fit of data in figure 7 from which we can infer an excellent linearity of the device responses with delivered dose.

| Description of sample | Sensitivity (nC/cGy) | Regression coefficient of the fit |
|---|---|---|
| 8.2 µm AZO electron selective contact. Sample 1 | 1.37 | 0.999880 |
| 8.2 µm AZO electron selective contact. Sample 2 | 1.48 | 0.999975 |
| 6.2 µm AZO electron selective contact. | 0.84 | 0.999897 |
| 8.2 µm TiO$_2$ electron selective contact. | 1.25 | 0.999918 |

Table 2. Sensitivity and linear regression coefficient of the various devices at 0V bias.

Table 2 demonstrates that the measured sensitivities are quite similar on the various 8.2 µm devices, and significantly lower for the 6.2 µm device as expected due to its reduced active layer thickness restricting the number of charges generated by the normally incident high energy photons.

The amorphous diodes show, in general, a sensitivity at 0V bias lower by a factor of 10 in respect to its crystalline silicon counterpart. Data reported by Mehran Yarahmadi et al. [5] and by Tomasz Koper et al. [6] show that a sensitivity of 8.5 nC/cGy is likely to be expected for a 10um of crystalline silicon diode at 50 kVp x-ray for a commercial device designed for dosimetry in radiotherapy.

Figure 8 shows the results of the measurement of devices at 30 V bias (20 V for the 6.2 µm device).

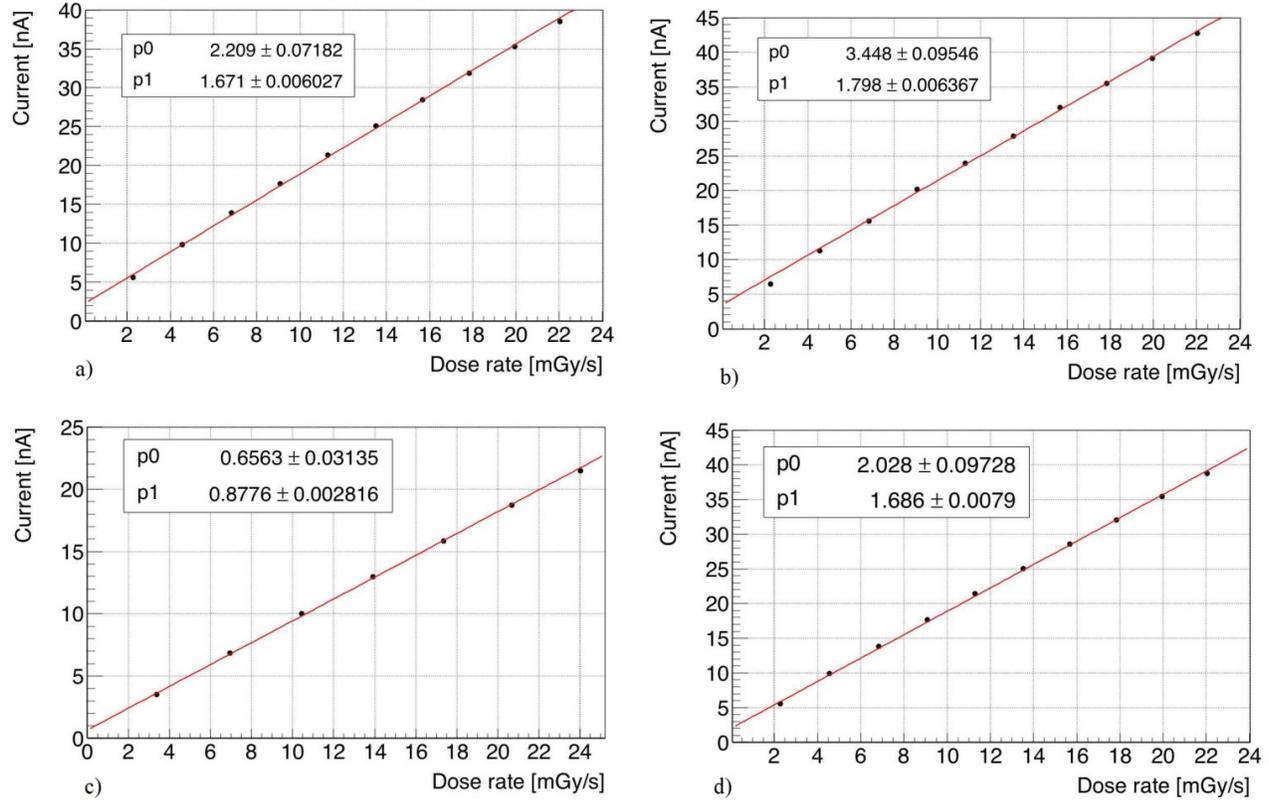

Fig.8 Current versus x-ray doses under 30 V bias of various devices in a-Si:H having different layer thickness and electron selective contact materials a) AZO electron selective contacts with a-Si:H layer having 8.2 µm thickness (Sample 1) b) AZO electron selective contacts with a-Si:H layer having 8.2 µm thickness (Sample 2) c) AZO electron selective contacts with a-Si:H layer having 6.2 µm thickness (biased at 20V) d) Titanium Oxide electron selective contacts with a-Si:H layer having 8.2 µm thickness.

Table 3 shows the calculated device sensitivities and the regression coefficient of the linear fitting applied to the results in Fig 8. Analogously with, the detector responses under a zero applied bias, we can again infer an excellent linearity of the response versus dose of the detectors when operated at an applied bias. Furthermore, the observed sensitivities for all the biased devices (30 V for 8.2 µm thick samples and 20 V for the 6.2 µm sample) show an increase by more than a factor 10 in comparison to device sensitivities at 0 V bias. While the linearities under non-zero bias are still excellent, it is worth noting the regression coefficients have slightly smaller values in devices under bias compared to device responses at 0 V bias as presented in Table 2. The higher sensitivity obtained under bias conditions shows a remarkable similarity with performance obtained with commercial crystalline silicon devices [5]

| Description of sample | Sensitivity (nC/cGy) | Regression coefficient of the fit |
|---|---|---|
| 8.2 µm AZO electron selective contact. Sample 1 | 16.71 | 0.999655 |
| 8.2 µm AZO electron selective contact. Sample 2 | 17.98 | 0.999491 |
| 6.2 µm AZO electron selective contact. | 8.78 | 0.999828 |
| 8.2 µm $TiO_2$ electron selective contact. | 16.86 | 0.999757 |

Table 3. Sensitivity and linear regression coefficient of the various devices under bias

## 4. Current versus voltage under x-ray irradiation.

Current versus voltage behavior of sample detectors under irradiation was also studies to further characterize them, with the X-ray tube configured as illustrated in figure 6 and held at fixed tube current and voltage (100 µA and 30 kV), producing a nominal dose rate of 6.8 mGy/s. The measured data (Fig.9) shows an exponential dependence of current from bias voltage according to the fit formula:

$$I = p_0 * (1 - p_1 * e^{-p_2 * V})$$

where I is the photocurrent, V is the bias voltage and p0, p1 and p2 are fitting parameters. Extrapolating the trend and assuming no breakdown at voltages above 40 V, the fit shows a saturating behavior of the current at a bias voltage in the order of ≈ $3/p_2$ (95% of saturation current) and a saturation photocurrent in the order of $p_0$. Hence, fitting the experimental data with the formula presented will allow for the determination of the saturation voltage and current of the devices. Table 4 shows the value of the saturation voltage and currents for all detectors. With the exception of the 8.2 µm AZO electron selective contact (Sample 2) which displays exceptionally high saturation values, the other 8.2 µm samples have similar values both in saturation current and voltage. The thinner 6.2 µm AZO sample has lower saturation voltage and current values as expected, with the reduction factor in saturation values in the 6.2 µm AZO sample (sample 4) in comparison to sample 1 (8.2 µm thick device with AZO electron selective contact) closely approximating the reduction factor in the active layer thickness between these two samples.

| Description of samples | Saturation Voltage [V] | Saturation Current [nA] |
|---|---|---|
| 8.2 µm AZO electron selective contact. Sample 1 | 209.2 | 38.48 |
| 8.2 µm AZO electron selective contact. Sample 2 | 333.4 | 64.44 |
| 6.2 µm AZO electron selective contact. | 144.2 | 22.74 |
| 8.2 µm TiO$_2$ electron selective contact. | 199.7 | 37.33 |

Table 4. Saturation voltages and saturation currents.

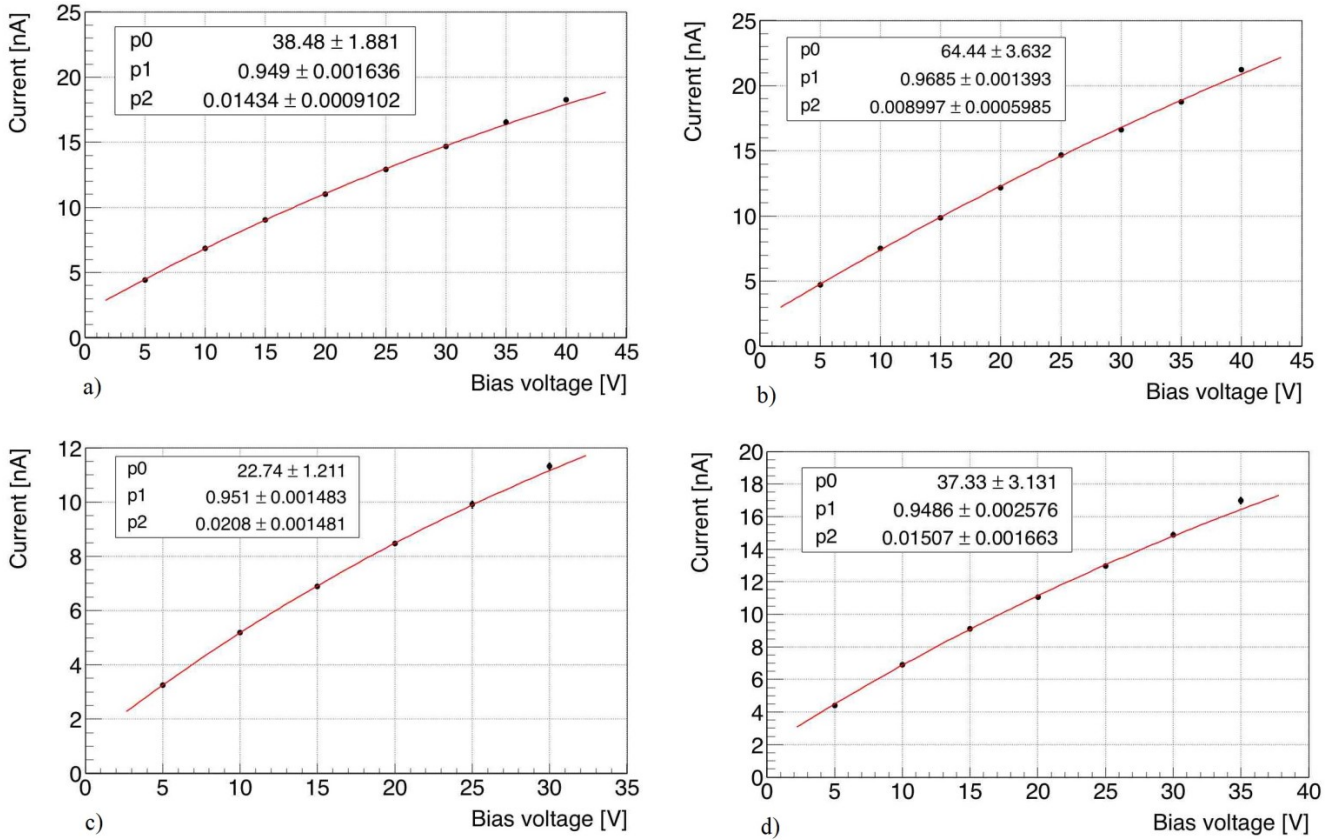

Fig.9 Current versus Bias Voltage a fixed x-ray tube settings (100 µA and 30 kV dose rate 6.8 mGy/s) of various devices in a-Si:H having different layer thickness and electron selective contact materials a) AZO electron selective contacts with a-Si:H layer having 8.2 µm thickness (Sample 1) b) AZO electron selective contacts with a-Si:H layer having 8.2 µm thickness (Sample 2) c) AZO electron selective contacts with a-Si:H layer having 6.2 µm thickness (biased at 20V) d) Titanium Oxide electron selective contacts with a-Si:H layer having 8.2 µm thickness.

## 5. Conclusions.

This work illustrates the performance of charge carrier selective contacts meant to be used for the fabrication of hydrogenated amorphous silicon diodes for x-ray detection applications using a 3D structure. For the first time, we explore the possibility to use charge selective contact aSi:H devices as particle detectors and, in order to assess their performance, experiments focused on leakage current, dose rate linearity and dose sensitivity measurements. Two types of electron selective contacts have been tested: Titanium ($TiO_2$) and Aluminum doped zinc oxide (AZO). In order to assess their usability for x-ray detection, two different thicknesses of the a-Si:H substrate have been adopted. The AZO appear to be the best candidate for all a-Si:H substrate thicknesses, showing a lower leakage current density between 9.2 and 14 nA/cm$^2$ in respect to the $TiO_2$ samples (over 24 nA/cm$^2$) for the device tested and reported in Table 1. The samples fabricated with the AZO contacts showed also a higher saturation voltage and current under illumination with 30 kVp x-rays. The variation of the leakage current with temperature confirms that mobility increases exponentially with temperature in a-Si:H as reported in literature [7] and AZO devices present a higher drift mobility than $TiO_2$ device. This result is confirmed also by the excellent linearity and sensitivity of the AZO samples, which peaked at a remarkable 18 nC/cGy. In photovoltaic mode, the sensitivity decreases to 1.8 nC/cGy which is significantly lower than crystalline silicon diodes commercially available. Despite this apparent disadvantage, the reproducibility and linearity shown by the amorphous diodes is comparable to commercial grade crystalline silicon devices. Thickness of the a-Si:H substrate is also another important parameter to consider in the optimization of the sensitivity of the diodes: a variation of the thickness from 6.2 to 8.2 um has generated an increment of sensitivity of more than 40% in the samples analyzed in this work. In conclusion, the Aluminum Zinc Oxide will be considered as the baseline combination for the fabrication of the 3D a-Si:H diodes.

## Acknowledgements

The research was funded by INFN Scientific committee 5 under the experiment name 3D-SiAm and by INFN committee for technological transfer under the experiment name INTEF-3D-SiAm. Additional funds came from Fondazione Cassa di Risparmio di Perugia (RISAI project n. 2019.0245). This work was also supported by the Australian Government Research Training Program, Australia, and the Australian Institute of Nuclear Science and Engineering Post-Graduate Research award, Australia (ALNSTU12583).